\documentstyle[twoside,fleqn,espcrc2]{article}

\newcommand{\AmS}{{\protect\the\textfont2
  A\kern-.1667em\lower.5ex\hbox{M}\kern-.125emS}}

\font\sevenrm=cmr7

\def\be{\begin{equation}}
\def\ee{\end{equation}}

\def\gsim{\mathrel{%
\rlap{\raise 0.511ex \hbox{$>$}}{\lower 0.511ex
\hbox{$\sim$}}}}

\def\lsim{\mathrel{
\rlap{\raise 0.511ex \hbox{$<$}}{\lower 0.511ex
\hbox{$\sim$}}}}

\def\up#1{\raise 1ex\hbox{\sevenrm#1}}

\def\build#1_#2^#3{\mathrel{
\mathop{\kern 0pt#1}\limits_{#2}^{#3}}}

\def\Lc{{\cal L}}

\font\tenbb=msbm10
\font\sevenbb=msbm7
\font\fivebb=msbm5
\newfam\bbfam
\textfont\bbfam=\tenbb \scriptfont\bbfam=\sevenbb
\scriptscriptfont\bbfam=\fivebb
\def\bb{\fam\bbfam}

\def\Rb{{\bb R}}

\title{Experimental Tests of Relativistic Gravity}

\author{Thibault DAMOUR\address{Institut des Hautes Etudes Scientifiques, 91440 
Bures-sur-Yvette, France\\
and DARC, CNRS - Observatoire de Paris, 92195 Meudon Cedex, France}}

\begin{document}

\begin{abstract}
The confrontation between Einstein's gravitation theory and experimental  
results, notably binary pulsar data, is summarized and its significance
discussed. Experiment and theory agree at the $10^{-3}$ level or better. All the 
basic structures of Einstein's theory (coupling of gravity to matter; 
propagation and self-interaction of the gravitational field, including in 
strong-field conditions) have been verified. However, the theoretical 
possibility that scalar couplings be naturally driven toward zero by the 
cosmological expansion suggests that the present agreement between Einstein's 
theory and experiment might be compatible with the existence of a long-range 
scalar contribution to gravity (such as the dilaton field, or a moduli field, of 
string theory). This provides a new theoretical paradigm, and new motivations 
for 
improving the experimental tests of gravity.
\end{abstract}

\maketitle

\section{Introduction}

Einstein's gravitation theory can be thought of as defined by two postulates.
One postulate states that the action functional describing the propagation and
self-interaction of the gravitational field is
\be
S_{\rm gravitation} = {c^4 \over 16\pi \ G} \int
{d^4 x \over c} \ \sqrt{g} \ R(g). \label{eq:01}
\ee
A second postulate states that the action functional describing the coupling of
all the fields describing matter and its electro-weak and strong interactions 
(leptons and quarks, gauge and Higgs bosons) is a (minimal) deformation of the 
special relativistic action functional used by particle physicists (the so 
called ``Standard Model''), obtained by replacing everywhere the flat Minkowski 
metric $\eta_{\mu \nu} = {\rm diag} (-1,+1,+1,+1)$ by $g_{\mu \nu} 
(x^{\lambda})$ 
and the partial derivatives $\partial_{\mu} \equiv \partial / \partial x^{\mu}$ 
by $g$-covariant derivatives $\nabla_{\mu}$. Schematically, one has
\be
S_{\rm matter} = \int { d^4 x \over c} \ \sqrt{g} \ \Lc_{\rm
matter} \ [\psi ,A_{\mu},H;g_{\mu \nu}]. \label{eq:02.a}  
\ee

Einstein's theory of gravitation is then defined by extremizing the total
action functional, $S_{\rm tot} \ [g,\psi ,A,H] = S_{\rm gravitation} \ [g] + 
S_{\rm matter} \ [\psi ,A,H,g]$.

Although, seen from a wider perspective, the two postulates (\ref{eq:01}) and
(\ref{eq:02.a}) follow from the unique requirement that the gravitational
interaction be mediated only by massless spin-2 excitations \cite{1}, the
decomposition in two postulates is convenient for discussing the theoretical
significance of various tests of General Relativity. Let us discuss in turn the
experimental tests of the coupling of matter to gravity (postulate 
(\ref{eq:02.a})), and the experimental tests of the dynamics of the 
gravitational field (postulate (\ref{eq:01})). For more details and references 
we refer the reader to \cite{2} or \cite{3}.

\section{Experimental tests of the coupling between matter and gravity}

The fact that the matter Lagrangian depends only on a symmetric
tensor $g_{\mu \nu} (x)$ and its first derivatives (i.e. the postulate of a 
universal ``metric coupling'' between matter and gravity) is a strong assumption 
(often referred to as the ``equivalence principle'') which has many observable
consequences for the behaviour of localized test systems embedded in given,
external gravitational fields. In particular, it predicts the constancy of the 
``constants'' (the outcome of local non-gravitational expe\-riments, referred to 
local standards, depends only on the values of the coupling constants and mass 
scales entering the Standard Model) and the universality of free fall (two test 
bodies dropped at the same location and with the same velocity in an
external gravitational field fall in the same way, independently of their masses
and compositions). 

Many sorts of data (from spectral lines in distant galaxies 
to a natural fission reactor phenomenon which took place at Oklo, Gabon, two 
billion years ago) have been used to set limits on a possible time variation of 
the basic coupling constants of the Standard Model. The best results concern the 
electromagnetic coupling, i.e. the fine-structure constant $\alpha_{\rm em}$. A 
recent reanalysis of the Oklo phenomenon gives a conservative upper bound 
\cite{5}
\be
-6.7 \times 10^{-17} \, {\rm yr}^{-1} < {{\dot{\alpha}}_{\rm em} \over
\alpha_{\rm em}} < 5.0 \times 10^{-17} \ {\rm yr}^{-1} , \label{eq:05} 
\ee
which is much smaller than the cosmological time scale $\sim 10^{-10} \
{\rm yr}^{-1}$. It would be interesting to confirm and/or improve the limit
(\ref{eq:05}) by direct laboratory measurements comparing clocks based on atomic
transitions having different dependences on $\alpha_{\rm em}$. [Current atomic
clock tests of the constancy of $\alpha_{\rm em}$ give the limit $\vert
{\dot{\alpha}}_{\rm em} / \alpha_{\rm em} \vert < 3.7 \times 10^{-14} \, {\rm
yr}^{-1}$ \cite{PTM95}.]

The universality of free fall has been verified at the $10^{-12}$ level both for
laboratory bodies \cite{7}, e.g. (from the last reference in \cite{7})
\be
\left( \frac{\Delta a}{a} \right)_{\rm Be \, Cu} = (-1.9 \pm 2.5) \times
10^{-12} \, , \label{eq:00.a}
\ee
and for the gravitational accelerations of the Moon and
the Earth toward the Sun \cite{8},
\be
\left(\frac{\Delta a}{a}\right)_{\rm Moon \, Earth} = (-3.2 \pm 4.6) \times
10^{-13} \, . \label{eq:00.b}
\ee

In conclusion, the main observable consequences of the Einsteinian postulate
(\ref{eq:02.a}) concerning the coupling between matter and gravity
(``equivalence principle'') have been verified with high precision by all
experiments to date (see Refs. \cite{2}, \cite{3} for discussions of other tests 
of the equivalence principle). The traditional paradigm (first put forward by 
Fierz \cite{10}) is that the extremely high precision of free fall experiments
($10^{-12}$ level) strongly suggests that the coupling between matter and
gravity is exactly of the ``metric'' form (\ref{eq:02.a}), but leaves open
possibilities more general than eq. (\ref{eq:01}) for the spin-content and
dynamics of the fields mediating the gravitational interaction. We shall
provisionally adopt this paradigm to discuss the tests of the other Einsteinian
postulate, eq. (\ref{eq:01}). However, we shall emphasize at the end that recent
theoretical findings suggest a new paradigm.

\section{Tests of the dynamics of the gravitational field in the weak
field regime}

Let us now consider the experimental tests of the dynamics of the gravitational
field, defined in General Relativity by the action functional (\ref{eq:01}).
Following first the traditional paradigm, it is convenient to enlarge our
framework by embedding General Relativity within the class of the most natural
relativistic theories of gravitation which satisfy exactly the matter-coupling
tests discussed above while differing in the description of the degrees of
freedom of the gravitational field. This class of theories are the
metrically-coupled tensor-scalar theories, first introduced by Fierz \cite{10}
in a work where he noticed that the class of non-metrically-coupled
tensor-scalar theories previously introduced by Jordan \cite{11} would
generically entail unacceptably large violations of the equivalence principle. 
The metrically-coupled (or equivalence-principle respecting) tensor-scalar 
theories are defined by keeping the postulate (\ref{eq:02.a}), but replacing the 
postulate (\ref{eq:01}) by demanding that the ``physical'' metric $g_{\mu \nu}$ 
(coupled to ordinary matter) be a composite object of the form 
\be
g_{\mu \nu} = A^2 (\varphi) \ g_{\mu \nu}^* , \label{eq:06}
\ee
where the dynamics of the ``Einstein'' metric $g_{\mu \nu}^*$ is defined by the
action functional (\ref{eq:01}) (written with the replacement $g_{\mu \nu}
\rightarrow g_{\mu \nu}^*$) and where $\varphi$ is a massless scalar field. 
[More generally, one can consider several massless scalar fields, with an action
functional of the form of a general nonlinear $\sigma$ model \cite{12}]. In 
other words, the action functional describing the dynamics of the spin 2 and 
spin 0 degrees of freedom contained in this generalized theory of gravitation 
reads
\begin{eqnarray}
S_{\rm gravitational} \, [g_{\mu \nu}^* ,\varphi ] 
\!\!\!\!\!\!&=\!\!\!\!\!& {c^4 \over 16\pi \, G_*} \int {d^4 x \over c} \, 
\sqrt{g_*} \nonumber \\ 
&\times\!\!\!\!\!&[R(g_*) - 2g_*^{\mu \nu} \, \partial_{\mu} \, \varphi \, 
\partial_{\nu} \, 
\varphi ] . \label{eq:07}
\end{eqnarray}
Here, $G_*$ denotes some bare gravitational coupling constant. This class of
theories contains an arbitrary function, the ``coupling function'' $A(\varphi)$.
When $A(\varphi) = {\rm const.}$, the scalar field is not coupled to matter and
one falls back (with suitable boundary conditions) on Einstein's theory. The
simple, one-parameter subclass $A(\varphi) = \exp (\alpha_0 \ \varphi)$ with
$\alpha_0 \in \Rb$ is the Jordan-Fierz-Brans-Dicke theory \cite{10},
\cite{J59}, \cite{BD}. In the general case, one can define the (field-dependent)
coupling strength of $\varphi$ to matter by 
\be
\alpha (\varphi) \equiv {\partial \ln A(\varphi) \over \partial
\varphi} . \label{eq:08}
\ee
It is possible to work out in detail the observable consequences of
tensor-scalar theories and to contrast them with the general relativistic case
(see, e.g., ref. \cite{12}).

Let us now consider the experimental tests of the dynamics of the
gravitational field that can be performed in the solar system. Because the
planets move with slow velocities $(v/c \sim 10^{-4})$ in a very weak
gravitational potential $(U/c^2 \sim (v/c)^2 \sim 10^{-8})$, solar system tests
allow us only to probe the quasi-static, weak-field regime of relativistic
gravity (technically  described by the so-called ``post-Newtonian'' expansion).
In the limit where one keeps only the first relativistic corrections to
Newton's gravity (first post-Newtonian approximation), all solar-system
gravitational experiments, interpreted within tensor-scalar theories, differ
from Einstein's predictions only through the appearance of two ``post-Einstein''
parameters $\overline{\gamma}$ and $\overline{\beta}$ (related to the usually
considered Eddington parameters $\gamma$ and $\beta$ through $\overline{\gamma}
\equiv \gamma -1$, $\overline{\beta} \equiv \beta -1$). The parameters
$\overline{\gamma}$ and $\overline{\beta}$ vanish in General Relativity, and are
given in tensor-scalar theories by  
\be
\overline{\gamma} = -2 \ {\alpha_0^2 \over 1+\alpha_0^2} , \label{eq:09.a}
\ee  
\be
\overline{\beta} = +{1 \over 2} \ {\beta_0 \ \alpha_0^2 \over
(1+\alpha_0^2)^2} , \label{ref:09.b}
\ee 
where $\alpha_0 \equiv \alpha (\varphi_0)$, $\beta_0 \equiv \partial \alpha
(\varphi_0) / \partial \varphi_0$; $\varphi_0$ denoting the
cosmologically-determined value of the scalar field far away from the solar
system. Essentially, the parameter $\overline{\gamma}$ depends only
on the linearized structure of the gravitational theory (and is a direct
measure of its field content, i.e. whether it is pure spin 2 or contains an
admixture of spin 0), while the parameter $\overline{\beta}$
parametrizes some of the quadratic nonlinearities in the field equations (cubic
vertex of the gravitational field). 

All currently performed gravitational experiments in the solar system, including 
perihelion advances of planetary orbits, the bending and delay of 
electromagnetic signals passing near the Sun, and very accurate range data to 
the Moon obtained by laser echoes, are compatible with the general relativistic 
predictions $\overline{\gamma} = 0 =\overline{\beta}$ and give upper bounds on 
both $\left\vert \overline{\gamma} \right\vert$ and $\left\vert \overline{\beta} 
\right\vert$(i.e. on possible fractional deviations from General Relativity). 
The best current limits come from: (i) VLBI measurements of the deflection of 
radio waves by the Sun, giving \cite{eubanks}: $-3.8 \times 10^{-4} < 
\overline{\gamma} < 2.6 \times 10^{-4}$, and (ii) Lunar Laser Ranging 
measurements of a possible polarization of the orbit of the Moon toward the Sun 
(``Nordtvedt effect'' \cite{N68}) giving \cite{8}: $4\overline{\beta} - 
\overline{\gamma} = -0.0007 \pm 0.0010$.

The corresponding bounds on the scalar coupling parameters $\alpha_0$ and 
$\beta_0$ are: $\alpha_0^2 < 1.9 \times 10^{-4}$, $-8.5 \times 10^{-4} < 
(1+\beta_0) \alpha_0^2 < 1.5 \times 10^{-4}$. Note that if one were working in 
the more general (and more plausible; see below) framework of theories where the 
scalar couplings violate the equivalence principle one would get much stronger 
constraints on the basic coupling parameter $\alpha_0$ of order $\alpha_0^2 
\lsim 
10^{-7}$ \cite{dv96a}.

The parametrization of the weak-field deviations between generic 
tensor-scalar theories and Einstein's theory
has been extended to the second post-Newto\-nian order \cite{14}. Only two
post-post-Einstein parameters, $\varepsilon$ and $\zeta$, representing a deeper
layer of structure of the gravitational interaction, show up. These parameters
have been shown to be already significantly constrained by binary-pulsar data:
$\vert \varepsilon \vert < 7 \times 10^{-2}$, $\vert \zeta \vert < 6 \times
10^{-3}$.

\section{Tests of the dynamics of the gravitational field in the
strong field regime}

In spite of the diversity, number and often high precision of solar system
tests, they have an important qualitative weakness : they probe neither the
radiation pro\-perties nor the strong-field aspects of relativistic gravity.
Fortunately, the discovery \cite{15} and continuous observational study of
pulsars in gravitationally bound binary orbits has opened up an entirely new
testing ground for relativistic gravity, giving us an experimental handle on the
regime of strong and/or radiative gravitational fields.

The fact that binary pulsar data allow one to probe the propagation properties
of the gravitational field is well known. This comes directly from the fact
that the finite velocity of propagation of the gravitational interaction
between the pulsar and its companion generates damping-like terms in the
equations of motion, i.e. terms which are directed against the velocities.
[This can be understood heuristically by considering that the finite velocity
of propagation must cause the gravitational force on the pulsar to make an
angle with the instantaneous position of the companion \cite{16}, and was
verified by a careful derivation of the general relativistic equations of motion
of binary systems of compact objects \cite{17}]. These damping forces cause the
binary orbit to shrink and its orbital period $P_b$ to decrease. The 
measurement, in some binary pulsar systems, of the secular orbital period decay 
$\dot{P}_b \equiv dP_b / dt$ \cite{18} thereby gives us a direct experimental 
probe of the damping terms present in the equations of motion.

The fact that binary pulsar data allow one to probe strong-field aspects of
re\-lativistic gravity is less well known. The a priori reason for saying that
they should is that the surface gravitational potential of a neutron star $Gm
/ c^2 R \simeq 0.2$ is a mere factor 2.5 below the black hole limit (and a
factor $\sim 10^8$ above the surface potential of the Earth). Due to the
peculiar ``effacement'' properties of strong-field effects taking place in
General Relativity \cite{17}, the fact that pulsar data probe the
strong-gravitational-field regime can only be seen when contrasting Einstein's
theory with more general theories. In particular, it has been found in
tensor-scalar theories \cite{19} that a self-gravity as strong as that of a
neutron star can naturally (i.e. without fine tuning of parameters) induce
order-unity deviations from general relativistic predictions in the orbital
dynamics of a binary pulsar thanks to the existence of nonperturbative
strong-field effects. [The adjective ``nonperturbative'' refers here to the fact 
that this phenomenon is nonanalytic in the coupling strength of the scalar 
field, eq. (\ref{eq:08}), which can be as small as wished in the weak-field 
limit]. As far as we know, this is the first example where large deviations from 
General Relativity, induced by strong self-gravity effects, occur in a theory 
which contains only positive energy excitations and whose post-Newtonian limit 
can be arbitrarily close to that of General Relativity.

A comprehensive account of the use of binary pulsars as laboratories for
testing strong-field gravity will be found in ref. \cite{20}. Two complementary
approaches can be pursued : a phenomenological one (``Parametrized
Post-Keplerian'' formalism), or a theory-dependent one \cite{12}, \cite{20}, 
\cite{22}.

The phenomenological analysis of binary pulsar timing data consists in fitting
the observed sequence of pulse arrival times to the generic DD timing formula
\cite{21} whose functional form has been shown to be common to the whole class 
of tensor-multi-scalar theories. The least-squares fit between the timing data
and the parameter-dependent DD timing formula allows one to measure, besides
some ``Keplerian'' parameters (``orbital period'' $P_b$, ``eccentricity''
$e$,$\ldots$), a maximum of eight ``post-Keplerian'' parameters: $k,\gamma
,\dot{P}_b ,r,s,\delta_{\theta} ,\dot e$ and $\dot x$. Here, $k\equiv
\dot{\omega} P_b / 2\pi$ is the fractional periastron advance per orbit,
$\gamma$ a time dilation parameter (not to be confused with its post-Newtonian
namesake), $\dot{P}_b$ the orbital period derivative mentioned above,
and $r$ and $s$ the ``range'' and ``shape'' parameters of the gravitational
(``Shapiro'') time delay caused by the companion. The important point is that 
the
post-Keplerian parameters can be measured without assuming any specific theory
of gravity. Now, each specific relativistic theory of gravity predicts that,
for instance, $k,\gamma, \dot{P}_b ,r$ and $s$ (to quote parameters that have
been successfully measured from some binary pulsar data) are some
theory-dependent functions of the (unknown) masses $m_1 ,m_2$ of the pulsar
and its companion. Therefore, in our example, the five simultaneous
phenomenological measurements of $k,\gamma ,\dot{P}_b ,r$ and $s$ determine,
for each given theory, five corresponding theory-dependent curves in the  $m_1
-m_2$ plane (through the 5 equations $k^{\rm measured} = k^{\rm theory} (m_1
,m_2 )$, etc$\ldots$). This yields three $(3=5-2)$ tests of the specified
theory, according to whether the five curves meet at one point in the mass
plane, as they should. [In the most general (and optimistic) case, discussed in
\cite{20}, one can phenomenologically analyze both timing data and
pulse-structure data (pulse shape and polarization) to extract up to nineteen
post-Keplerian parameters.] The theoretical significance of these tests depends 
upon the physics lying behind the post-Keplerian parameters involved in the 
tests. For instance, as we said above, a test involving $\dot{P}_b$ probes the 
propagation (and helicity) properties of the gravitational interaction. But a 
test involving, say, $k,\gamma ,r$ or $s$ probes (as shown by combining the 
results of \cite{12} and \cite{19}) strong self-gravity effects independently of 
radiative effects.

Besides the phenomenological analysis of binary pulsar data, one can also
adopt a theory-dependent methodology \cite{12}, \cite{20}, \cite{22}. The idea 
here is to work from the start within a certain finite-dimensional ``space of 
theories'', i.e. within a specific class of gravitational theories labelled by 
some theory parameters. Then by fitting the raw pulsar data to the predictions 
of the considered class of theories, one can determine which regions of 
theory-space are compatible (at say the 90\% confidence level) with the 
available experimental data. This method can be viewed as a strong-field 
genera\-lization of the parametrized post-Newtonian formalism \cite{2} used to 
analyze solar-system experiments. When non-perturbative strong-field effects are 
absent one can parametrize strong-gravity effects in neutron stars by using an 
expansion in powers of the ``compactness'' $c_A \equiv -2 \ \partial \ {\rm ln} 
\ m_A / \partial \ {\rm ln} \ G \sim G \ m_A / c^2 \ R_A$. Ref. \cite{12} has 
then shown that the observable predictions of generic tensor-multi-scalar 
theories could be parametrized by a sequence of ``theory parameters'', 
$\overline{\gamma} \ , \ \overline{\beta} \ , \ \beta_2 \ , \ \beta' \ , \ 
\beta'' \ , \ \beta_3 \ , \ (\beta \beta') \ldots$ representing deeper and 
deeper layers of structure of the relativistic gravitational interaction beyond 
the first-order post-Newtonian level parametrized by $\overline{\gamma}$ and 
$\overline{\beta}$. When non-perturbative strong-field effects develop, one 
cannot use the multi-parameter approach just mentioned. A useful alternative 
approach is then to work within specific, low-dimensional ``mini-spaces of 
theories''. Of particular interest is the two-dimensional mini-space of 
tensor-scalar theories defined by the coupling function $A(\varphi) = {\rm exp} 
\left( \alpha_0 \, \varphi + {1\over 2} \, \beta_0 \, \varphi^2 \right)$. The 
predictions of this family of theories (parametrized by $\alpha_0$ and 
$\beta_0$) are analytically described, in weak-field contexts, by the 
post-Einstein parameter (\ref{eq:09.a}), and can be studied in strong-field 
contexts by combining analytical and numerical methods \cite{22}.

Let us now briefly summarize the current experimental situation. Concerning the 
first discovered binary pulsar PSR$1913+16$ \cite{15}, it has been possible to 
measure with accuracy the three post-Keplerian para\-meters $k, \gamma$ and 
$\dot{P}_b$. From what was said above, these three simultaneous measurements 
yield {\it one} test of gravitation theories. After subtracting a small ($\sim 
10^{-14}$ level in $\dot{P}_b$ !), but significant, perturbing effect caused by 
the Galaxy \cite{23}, one finds that General Relativity passes this $( k-\gamma
-\dot{P}_b )_{1913+16}$ test with complete success at the $10^{-3}$ level. More 
precisely, one finds \cite{24}, \cite{18}
\begin{eqnarray}
\left[ \frac{{\dot{P}}_b^{\rm obs} - {\dot{P}}_b^{\rm
galactic}}{{\dot{P}}_b^{\rm GR} [k^{\rm obs} ,\gamma^{\rm
obs}]}\right]_{1913+16} \!\!\!\!\!\!\!\!&=\!\!\!\!\!& 1.0032 \pm 0.0023 ({\rm 
obs})\nonumber \\
\!\!\!\!\!\!\!\!&\pm \!\!\!\!\!& 0.0026 ({\rm galactic}) \nonumber \\
\!\!\!\!\!\!\!\!&=\!\!\!\!\!& 1.0032 \pm 0.0035 \, , \label{eq:00g}
\end{eqnarray}
where ${\dot{P}}_b^{\rm GR} [k^{\rm obs} ,\gamma^{\rm obs}]$ is the GR
prediction for the orbital period decay computed from the observed values of
the other two post-Keplerian parameters $k$ and $\gamma$.

This beautiful confirmation of General Relativity is
an embarrassment of riches in that it probes, at the same time, the propagation
{\it and} strong-field properties of relativistic gravity ! If the timing
accuracy of PSR$1913+16$ could improve by a significant factor two more
post-Keplerian parameters ($r$ and $s$) would become measurable and would allow
one to probe separately the propagation and strong-field aspects \cite{24}.
Fortunately, the discovery of the binary pulsar PSR$1534+12$ \cite{25} (which
is significantly stronger than PSR$1913+16$ and has a more favourably oriented
orbit) has opened a new testing ground, in which it has been possible to probe
strong-field gravity independently of radiative effects. A phenomenological
analysis of the timing data of PSR$1534+12$ has allowed one to measure the four
post-Keplerian parameters $k,\gamma ,r$ and $s$ \cite{24}. From what was said
above, these four simultaneous measurements yield {\it two} tests of 
strong-field gravity, without mixing of radiative effects. General Relativity is 
found to pass these tests with complete success within the measurement accuracy
\cite{24}, \cite{18}. The most precise of these new, pure strong-field tests is
the one obtained by combining the measurements of $k$, $\gamma$ and $s$. Using
the most recent data \cite{stairs} one finds agreement at the 1\% level:
\be
\left[ \frac{s^{\rm obs}}{s^{\rm GR} [k^{\rm obs} ,\gamma^{\rm
obs}]}\right]_{1534+12} = 1.007 \pm 0.008 \, . \label{eq:00h} 
\ee
Recently, it has been possible to extract also the ``radiative''
parameter $\dot{P}_b$ from the timing data of PSR$1534+12$. Again, General
Relativity is found to be fully consistent (at the $\sim 15\%$ level)
with the additional test provided by the $\dot{P}_b$ measurement
\cite{stairs}. Note that this gives our second direct experimental confirmation 
that the gravitational interaction propagates as predicted by Einstein's theory.

More recently, measurements of the pulse shape of PSR $1913+16$ \cite{kramer}, 
\cite{taylorweisberg} have detected a time variation of the pulse shape 
compatible with the prediction \cite{dr74}, \cite{boc} that the general 
relativistic spin-orbit coupling should cause a secular change in the 
orientation of the pulsar beam with respect to the line of sight (``geodetic 
precession''). As envisaged long ago \cite{dr74} this precession will cause the 
pulsar to disappear (around 2035) and to remain invisible for hundreds of years 
\cite{kramer}, \cite{taylorweisberg}.

A theory-dependent analysis of the published pulsar data on PSRs $1913+16$, 
$1534+12$ and $0655+64$ (a dissymetric system constraining the existence of 
dipolar radiation \cite{WZ89}) has been recently performed within the $(\alpha_0 
, \beta_0)$-space of tensor-scalar theories introduced above \cite{22}. This 
analysis proves that binary-pulsar data exclude large regions of theory-space 
which are compatible with solar-system experiments. This is illustrated in Fig. 
9 of Ref.~\cite{22} which shows that $\beta_0$ must be larger than about $-5$, 
while any value of $\beta_0$ is compatible with weak-field tests as long as 
$\alpha_0$ is small enough.

\section{Was Einstein 100\% right ?}

Summarizing the experimental evidence discussed above, we can say that
Einstein's postulate of a pure metric coupling between matter and gravity
(``equivalence principle'') appears to be, at least, $99.999 \ \! 999 \ \! 999 \
\! 9\%$ right (because of universality-of-free-fall experiments), while
Einstein's postulate (\ref{eq:01}) for the field content and dynamics of the
gravitational field appears to be, at least, $99.9\%$ correct both in the
quasi-static-weak-field limit appropriate to solar-system experiments, and in
the radiative-strong-field regime explored by binary pulsar experiments. Should
one apply Occam's razor and decide that Einstein must have been $100\%$ right,
and then stop testing General Relativity ? My answer is definitely, no !

First, one should continue testing a basic physical theory such as Ge\-neral
Relativity to the utmost precision available simply because it is one of the
essential pillars of the framework of physics. This is the fundamental
justification of an experiment such as Gravity Probe B (the Stanford gyroscope
experiment), which will advance by one order of magnitude our experimental
knowledge of post-Newtonian gravity. 

Second, some very crucial qualitative features of General Relativity have not 
yet been verified : in particular the existence of black holes, and the direct 
detection on Earth of gravitational waves. Hopefully, the LIGO/VIRGO network of 
interferometric detectors will observe gravitational waves early in the next 
century.

Last, some theoretical findings suggest that the current level of
precision of the experimental tests of gravity might be naturally (i.e.
without fine tu\-ning of parameters) compatible with Einstein being actually
only 50\% right ! By this we mean that the correct theory
of gravity could involve, on the same fundamental level as the Einsteinian
tensor field $g_{\mu \nu}^*$, a massless scalar field $\varphi$.

Let us first question the traditional paradigm \cite{10}, \cite{2} according to 
which special attention should be given to tensor-scalar theories respecting the
equivalence principle. This class of theories was, in fact, introduced in a
purely {\it ad hoc} way so as to prevent too violent a contradiction with
experiment. However, it is important to notice that the scalar couplings which
arise naturally in theories unifying gravity with the other interactions
systematically violate the equivalence principle. This is true both in
Kaluza-Klein theories (which were the starting point of Jordan's theory) and in
string theories. In particular, it is striking that (as first noted by Scherk 
and Schwarz \cite{SS74}) the dilaton field $\Phi$, which plays an essential role 
in string theory, appears as a necessary partner of the graviton field $g_{\mu 
\nu}$ in all string models. Let us recall that $g_s = e^{\Phi}$ is the basic 
string coupling constant (measuring the weight of successive string loop 
contributions) which determines, together with other scalar fields (the moduli), 
the values of all the coupling constants of the low-energy world. This means, 
for instance, that the fine-structure constant $\alpha_{\rm em}$ is a function 
of $\Phi$ (and possibly of other moduli fields). In intuitive terms, while 
Einstein proposed a framework where geometry and gravitation were united as a 
dynamical field $g_{\mu \nu} (x)$, i.e. a soft structure influenced by the 
presence of matter, string theory extends this idea by proposing a framework 
where geometry, gravitation, gauge couplings and gravitational couplings all 
become soft structures described by interrelated dynamical fields. Symbolically, 
one has $g_{\mu \nu} (x) \sim g^2 (x) \sim G(x)$. This spatiotemporal 
variability of coupling constants entails a clear violation of the equivalence 
principle. In particular, $\alpha_{\rm em}$ would be expected to vary on the 
Hubble time scale (in contradiction with the limit (\ref{eq:05}) above), and 
materials of different compositions would be expected to fall with different 
accelerations (in contradiction with the limits (\ref{eq:00.a}), (\ref{eq:00.b}) 
above).

The most popular idea for reconciling gravitational experiments with the
existence, at a fundamental level, of scalar partners of $g_{\mu \nu}$ is to
assume that all these scalar fields (which are massless before supersymmetry
breaking) will acquire a mass after supersymmetry breaking. Typically one 
expects this mass $m$ to be in the TeV range \cite{CCQR}. This would ensure that 
scalar exchange brings only negligible, exponentially small corrections $\propto 
\ \exp (-mr/\hbar c)$ to the general relativistic predictions concerning 
low-energy gravitational effects. However, the interesting possibility exists 
that the mass $m$ be in the milli eV range, corresponding to observable 
deviations 
from usual gravity below one millimeter \cite{ia}, \cite{fkz}, \cite{dimo}.

But, the idea of endowing the scalar partners of $g_{\mu \nu}$ with a non zero 
mass is fraught with many cosmological difficulties \cite{BS93}, \cite{29}, 
\cite{DV96b}. Though these cosmological difficulties might be solved by a 
combination of ad hoc solutions (e.g. introducing a secondary stage of inflation 
to dilute previously produced dilatons \cite{RT95}, \cite{LS95}), a more radical 
solution to the problem of reconciling the existence of the dilaton (or any 
moduli field) with experimental tests and cosmological data has been proposed 
\cite{30} (see also \cite{28} which considered an 
equivalence-principle-respecting scalar field). The main idea of Ref. \cite{30} 
is that string-loop effects (i.e. corrections depending upon $g_s = e^{\Phi}$ 
induced by worldsheets of arbitrary genus in intermediate string states) may 
modify the low-energy, Kaluza-Klein type matter couplings $(\propto \, e^{-2 
\Phi} \, F_{\mu \nu} \, F^{\mu \nu})$ of the dilaton (or moduli) in such a 
manner that the VEV of $\Phi$ be cosmologically driven toward a finite value 
$\Phi_m$ where it decouples from matter. For such a ``least coupling principle'' 
to hold, the loop-modified coupling functions of the dilaton, $B_i (\Phi) = 
e^{-2\Phi} + c_0 +c_1 \, e^{2\Phi} +\cdots +$ (nonperturbative terms), must 
exhibit extrema for finite values of $\Phi$, and these extrema must have certain 
universality properties. A natural way in which the required conditions could 
be satisfied is through the existence of a discrete symmetry in scalar space. 
[For instance, a symmetry under $\Phi \rightarrow -\Phi$ would guarantee that 
all the scalar coupling functions reach an extremum at the self-dual point 
$\Phi_m =0$]. 

A study of the efficiency of this mechanism of cosmological attraction of 
$\varphi$ towards 
$\varphi_m$ ($\varphi$ denoting the canonically normalized scalar field in the 
Einstein frame, see Eq.~(\ref{eq:07})) estimates that the present vacuum 
expectation value $\varphi_0$ of the scalar field would differ (in a rms sense) 
from $\varphi_m$ by   
\be
\varphi_0 - \varphi_m \sim 2.75 \times 10^{-9} \times \kappa^{-3} \,
\Omega_m^{-3/4} \, \Delta \varphi \, . \label{eq:12}
\ee
Here $\kappa$ denotes the curvature of the gauge coupling function ${\rm ln} \ 
B_F (\varphi)$ around the maximum $\varphi_m$, $\Omega_m$ denotes the present 
cosmological matter density in units of $10^{-29} \, g$ cm$^{-3}$, and $\Delta 
\varphi$ the deviation 
$\varphi - \varphi_m$ at the beginning of the (classical) radiation era. 
Equation (\ref{eq:12}) predicts (when $\Delta \varphi$ is of order 
unity\footnote{However, $\Delta \varphi$ could be $\ll 1$ if the attractor 
mechanism already applies during an early stage of potential-driven inflation 
\cite{31}.}) the existence, at the present cosmological epoch, of many small, 
but not unmeasurably small, deviations from General Relativity proportional to 
the {\it square} of $\varphi_0 -\varphi_m$. This provides a new incentive for 
trying to improve by several orders of magnitude the various experimental tests 
of Einstein's equivalence principle. The most sensitive way to look for a small 
residual violation of the equivalence principle is to perform improved tests of 
the universality of free fall. The mechanism of Ref. \cite{30} suggests a 
specific composition-dependence of the residual differential acceleration of 
free fall and estimates that a non-zero signal could exist at the very small 
level 
\be 
\left( {\Delta a \over a}
\right)_{\rm rms}^{\rm max} \sim 1.36 \times 10^{-18} \, \kappa^{-4} \,
\Omega_m^{-3/2} \, (\Delta \varphi)^2 , \label{eq:13} 
\ee
where $\kappa$ is expected to be of order unity (or smaller, leading to a
larger signal, in the case where $\varphi$ is a modulus rather than the
dilaton). 

Let us emphasize that the strength of the cosmological scenario
considered here as counterargument to applying Occam's razor lies in the fact
that the very small number on the right-hand side of eq. (\ref{eq:13}) has been
derived without any fine tuning or use of small parameters, and turns out to be
naturally smaller than the $10^{-12}$ level presently tested by
equivalence-principle experiments (see equations (\ref{eq:00.a}),
(\ref{eq:00.b})). The estimate (\ref{eq:13}) gives added significance to the
project of a Satellite Test of the Equivalence Principle (nicknamed STEP, and
currently studied by NASA, ESA and CNES) which aims at probing the universality 
of free fall of pairs of test masses orbiting the Earth at the
$10^{-18}$ level \cite{32}.

\end{document}